\documentclass[epj]{webofc}
\usepackage[utf8]{inputenc}
\usepackage[varg]{txfonts}   
\usepackage{booktabs}
\usepackage{xcolor}
\definecolor{darkred}{rgb}{0.4,0.0,0.0}
\definecolor{darkgreen}{rgb}{0.0,0.4,0.0}
\definecolor{darkblue}{rgb}{0.0,0.0,0.4}
\usepackage[bookmarks,linktocpage,colorlinks,
    linkcolor = darkred,
    urlcolor  = darkblue,
    citecolor = darkgreen]{hyperref}
\wocname{EPJ Web of Conferences}
\woctitle{Lattice2017}
\graphicspath{{figs/}}
\newcommand{\lr}[1]{ \left( #1 \right) }
\newcommand{\beq}{\begin{equation}}
\newcommand{\eeq}{\end{equation}\noindent}
\newcommand{\TR}{\mbox{Tr}}
\newcommand{\mDU}{\mathcal{D}U}

\begin{document}
%
\selectlanguage{english}
\title{%
Restoring canonical partition functions from imaginary chemical potential
}
\author{%
\firstname{V.G.}  \lastname{Bornyakov}\inst{1,2}
\and
\firstname{D.}  \lastname{Boyda}\inst{1,3}
\and
\firstname{V.}  \lastname{Goy}\inst{1}\fnsep\thanks{Speaker, \email{vovagoy@gmail.com}}
\and
\firstname{A.}  \lastname{Molochkov}\inst{1}
\and
\firstname{A.}  \lastname{Nakamura}\inst{1,5,6}
\and
\firstname{A.}  \lastname{Nikolaev}\inst{1}
\and
\firstname{V.I.}  \lastname{Zakharov}\inst{1,4,7}
}
\institute{%
School of Biomedicine, Far Eastern Federal University, Sukhanova 8, 690950 Vladivostok, Russia
\and
Institute for High Energy Physics NRC Kurchatov Institute, 142281 Protvino, Russia
\and
School of Natural Sciences, Far Eastern Federal University, Sukhanova 8, 690950 Vladivostok, Russia
\and
Institute of Theoretical and Experimental Physics NRC Kurchatov Institute, 117218 Moscow, Russia
\and
Theoretical Research Division, Nishina Center, RIKEN, Wako 351-0198, Japan
\and
Research Center for Nuclear Physics (RCNP), Osaka University, Ibaraki, Osaka, 567-0047, Japan
\and
Moscow Institute of Physics and Technology, Dolgoprudny, Moscow Region, 141700 Russia
}
\abstract{%
  Using GPGPU techniques and multi-precision calculation we developed the code to study QCD phase transition line in the canonical approach. The canonical approach is a powerful tool to investigate sign problem in Lattice QCD. The central part of the canonical approach is the fugacity expansion of the grand canonical partition functions. Canonical partition functions $Z_n(T)$ are coefficients of this expansion. Using various methods we study properties of $Z_n(T)$. At the last step we perform cubic spline for temperature dependence of $Z_n(T)$ at fixed $n$ and compute baryon number susceptibility $\chi_B/T^2$ as function of temperature. After that we compute numerically $\partial\chi/ \partial T$ and restore crossover line in QCD phase diagram. We use improved Wilson fermions and Iwasaki gauge action on the $16^3 \times 4$ lattice with $m_{\pi}/m_{\rho} = 0.8$ as a sandbox to check the canonical approach. In this framework we obtain coefficient in parametrization of crossover line $T_c(\mu_B^2)=T_c\lr{c-\kappa\, \mu_B^2/T_c^2}$ with $\kappa = -0.0453 \pm 0.0099$.
}
\maketitle
\section{Introduction}\label{intro}
  A lattice QCD simulation is a first-principles calculation, and this
  makes it possible to study the quark/hadron world using a non-perturbative approach.
  The basic formula is the path integral form of the grand canonical
  partition function:
  \beq
    Z_G(\mu,T) = \TR\, e^{-(\hat{H}-\mu \hat{N})/T} 
    = \int \mDU (\det\Delta(\mu))^{N_f} e^{-S_G},
  \eeq
  where $\mu$ is the chemical potential, $T$ is
  the temperature, $\hat{H}$ is the Hamiltonian, $\hat{N}$ is the quark number operator, $\det\Delta(\mu)$ is the fermion determinant,
  and $S_G$ is the gluon field action.
  We use $\mu$ notation for quark chemical potential and $\mu_B$ for baryon chemical potential.
  In this paper, we consider the two-flavor case: $N_f=2$.

  To explore the finite density QCD, we consider finite
  $\mu$ regions.  However, when $\mu$ takes a nonzero real value,
  the fermion determinant becomes a complex number. 
  This is problematic, because in the Monte Carlo simulations,
  we generate the gluon fields with the probability
  \beq
    P = (\det\Delta(\mu))^{N_f} e^{-S_G}/Z,
  \eeq
  and if the fermion determinant is complex, we are in trouble.
  In principle, we may write $\det\Delta=|\det\Delta|\exp(\imath \phi)$,
  perform the Monte Carlo update with $|\det\Delta|$, and 
  push the phase $\exp(\imath \phi)$ into an observable. 
  In the lowest order of $\mu$, the phase is given by
  $\phi=N_f\, \mu\, \mathrm{Im}\, \TR D^{-1} \partial D/\partial\mu$
  \cite{PhysRevD.66.074507}, which is proportional to the volume. 
  Therefore, the phase fluctuations grow with the volume,
  so this does not work in practice.

  Recently we presented~\cite{Bornyakov:2016wld,Boyda:2017lps,Goy:2016egl}
  new method within the canonical approach, or the fugacity expansion~\cite{Borici:2004bq,
  Alexandru:2005ix,deForcrand:2006ec,nagata2010wilson,Alexandru:2010yb,
  li2010finite,Li:2011ee,Nagata:2012tc,Danzer:2012vw,Gattringer:2014hra},
  which is a candidate for solving the sign problem.
  In the canonical approach, the grand canonical partition function is
  expressed as a fugacity expansion:
  \beq
    Z_G(\mu,T) = \sum_{n=-\infty}^{+\infty} Z_n(T) \xi^n,
    \label{Eq:FugacityExpansion}
  \eeq
  where
  $\xi = \exp(\mu/T)$ is the fugacity.
  Both $Z_G$ and $Z_n$ are functions of the volume $V$ which we omit in the arguments.

  The inverse transformation~\cite{hasenfratz1992canonical} is
  \beq
    Z_n(T) = \int_0^{2\pi}\frac{d\phi}{2\pi} e^{-\imath n\phi} Z_G(\xi=e^{\imath\phi},T).
    \label{Eq:InverseFugacityExpansion}
  \eeq

  In Eq.\ (\ref{Eq:FugacityExpansion}), 
  canonical partition functions $Z_n$ do not depend on $\mu$,
  and Eq.\ (\ref{Eq:FugacityExpansion}) works for real, 
  imaginary, and even complex $\mu$.
  When the chemical potential is pure imaginary, $\mu=i\mu_I$,
  the fermion determinant is real, and in those regions,
  we can construct $Z_n$ from $Z_G$.  
  After determining $Z_n$ in this way, we can
  study real physical $\mu$ regions using formula 
  (\ref{Eq:FugacityExpansion}).
  
  \begin{figure}[h]
    \begin{minipage}[t]{1.\textwidth}
      \centering
      \includegraphics[width=.5\textwidth]{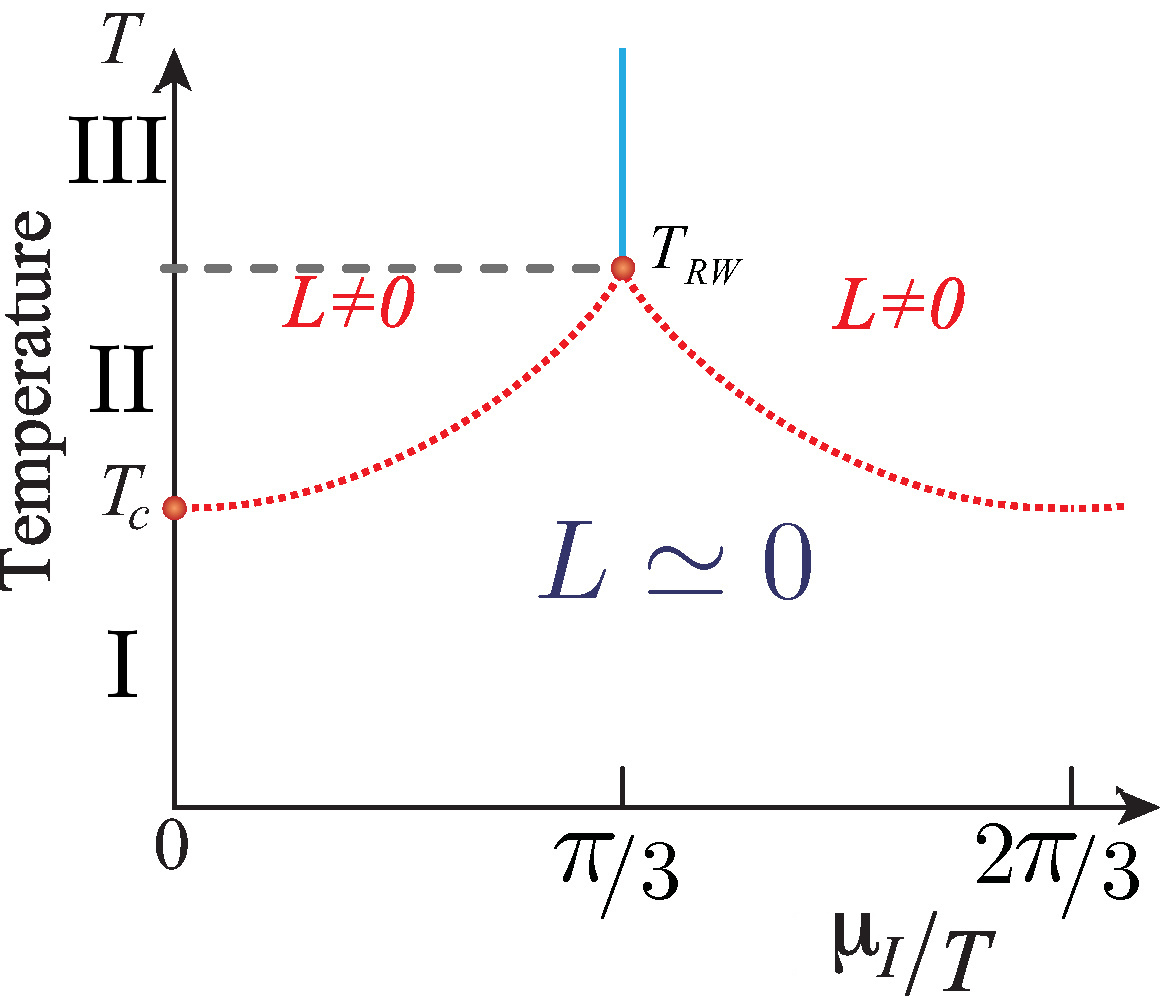}
      \caption{Three regimes for imaginary $\mu_I$. \label{fig:RWphase}}
    \end{minipage}
  \end{figure}
  
\section{Calculation of $Z_n$}

  We use improved Wilson fermions and Iwasaki gauge action on the $16^3 \times 4$ lattice with $m_{\pi}/m_{\rho} = 0.8$.
  All parameters of this action we take from WHOT-QCD Collaboration~\cite{Ejiri:2009hq}.
  In this framework we compute quark number density at seven values of temperature.
  Then we can restore grand canonical partition function by the integration of the density $n(\mu_I)$
  \beq
    \frac{Z_G(\mu_I, T)}{Z_G(0, T)} =
    \exp\lr{-V \int_0^{\mu_I/T}  \mathrm{d}x \, n(x)}\,.
    \label{eq:ZGbyIntegrate}
  \eeq
 
\begin{figure}[h]
    \begin{minipage}[t]{0.48\textwidth}
      \centering
      \includegraphics[width=1.\textwidth]{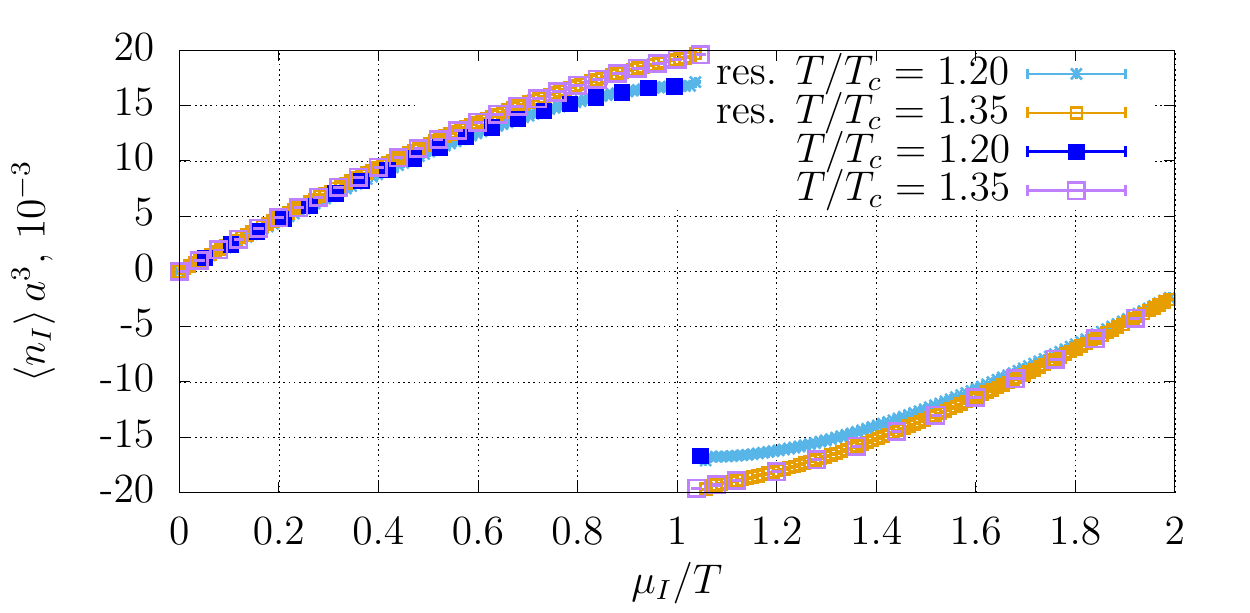}
      
      \vspace{-1.5ex}
      \includegraphics[width=1.\textwidth]{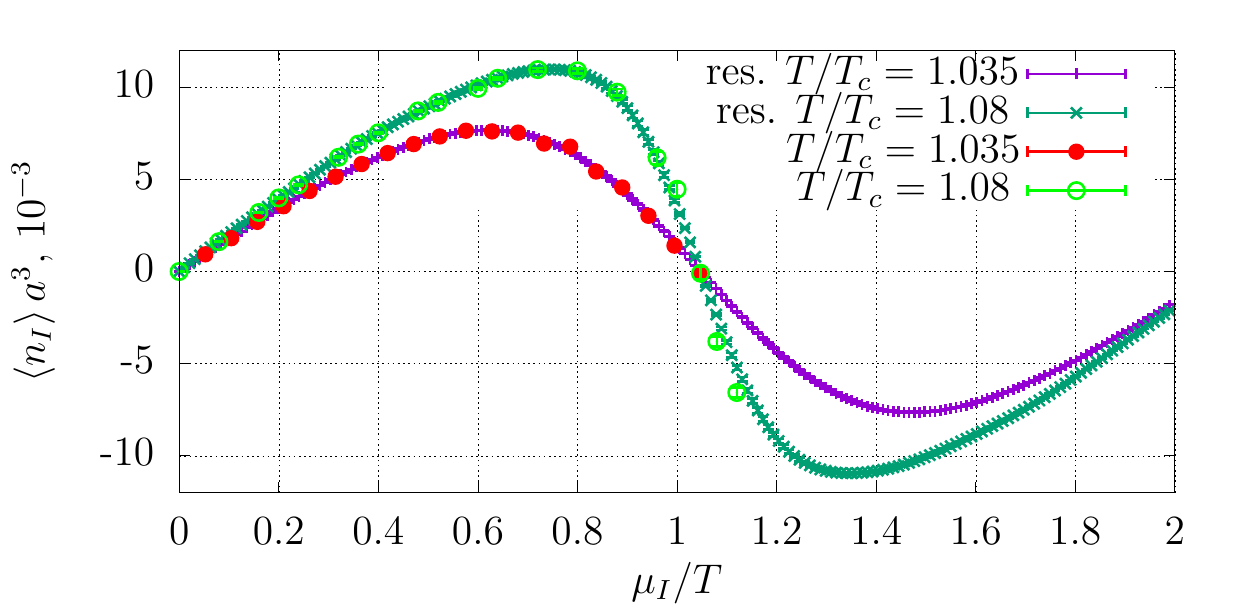}
      
      \vspace{-1.5ex}
      \includegraphics[width=1.\textwidth]{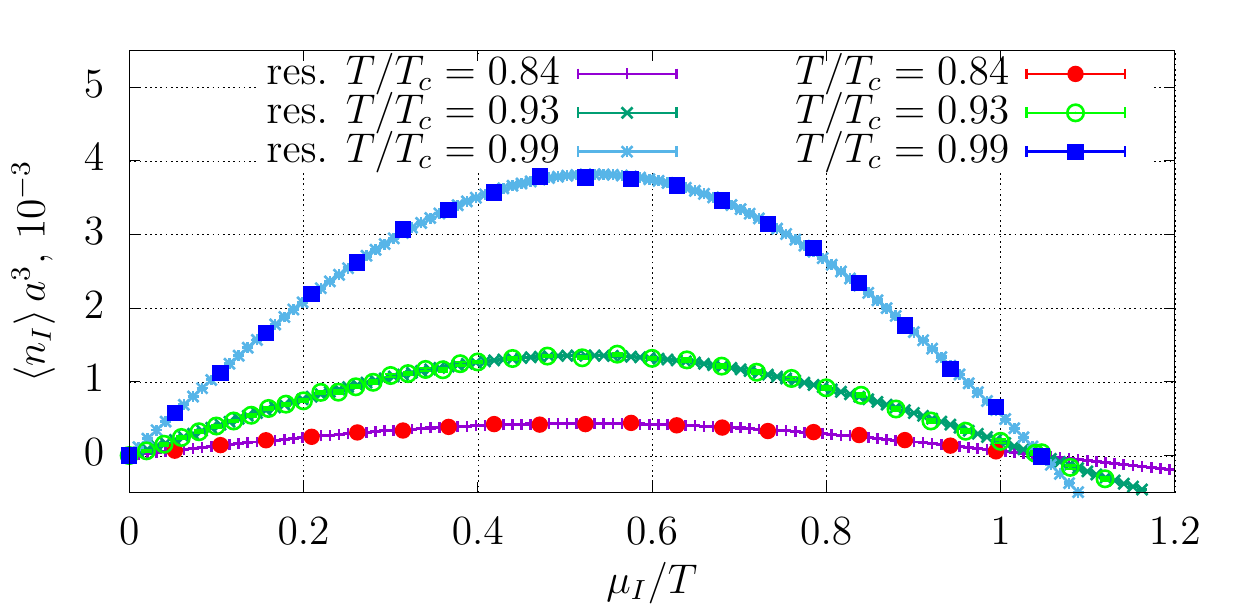}
    
      \caption{Baryon density for imaginary $\mu_I$ (res. -- restored from $Z_n$). \label{fig:bdi}}
    \end{minipage}
    \hspace{\fill}
    \begin{minipage}[t]{0.48\textwidth}
      \centering
      \includegraphics[width=1.\textwidth]{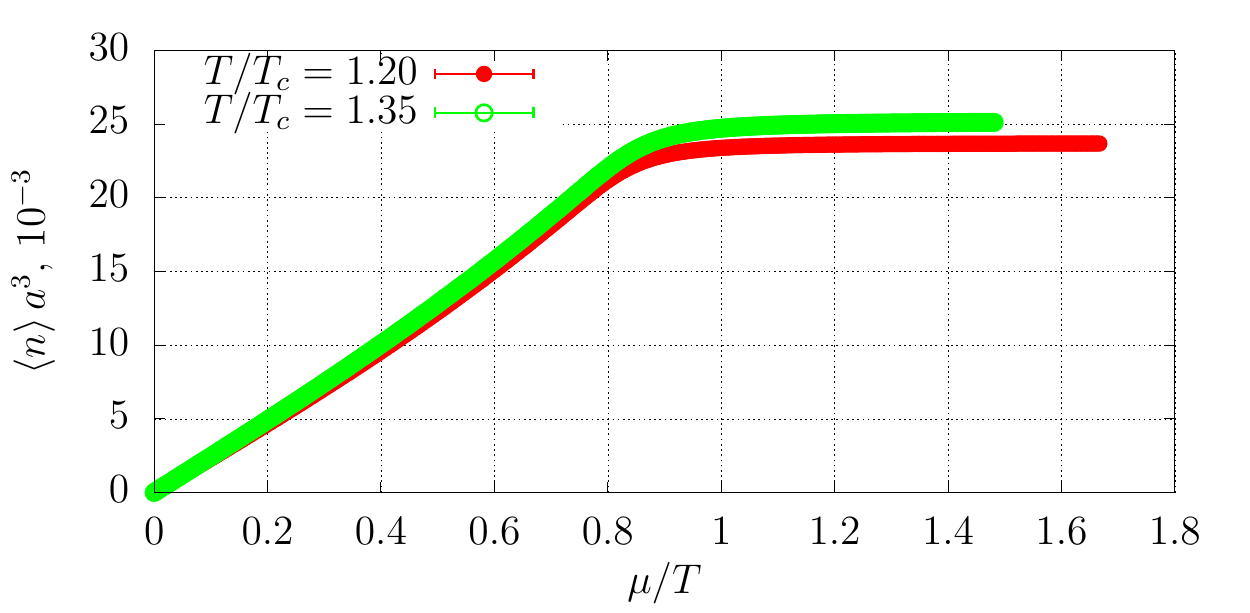}
      
      \vspace{-1.5ex}
      \includegraphics[width=1.\textwidth]{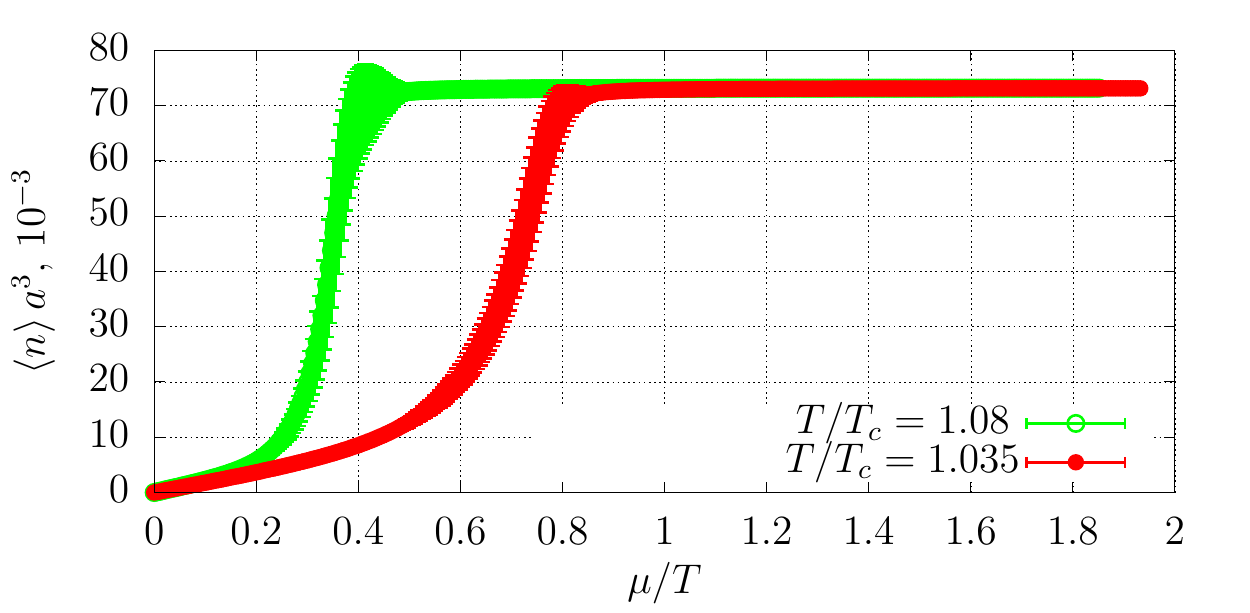}
      
      \vspace{-1.5ex}
      \includegraphics[width=1.\textwidth]{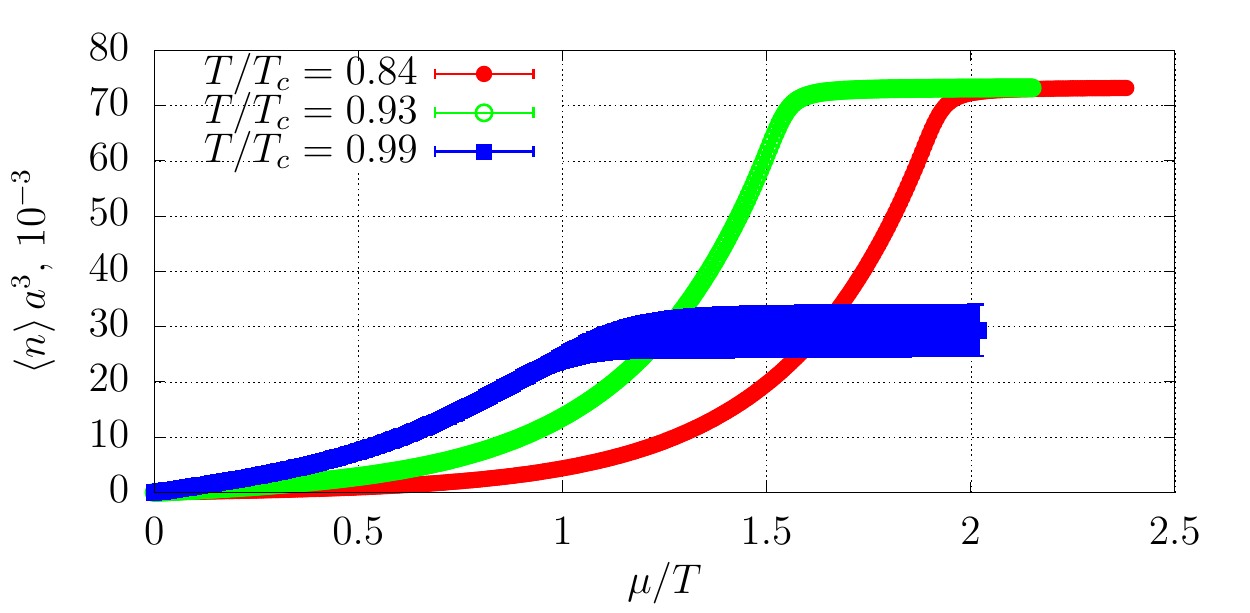}
    
      \caption{Baryon density for real quark chemical potential $\mu$. \label{fig:bdr}}
    \end{minipage}
  \end{figure}

  Phase diagram for imaginary value of chemical potential has non trivial structure.
  As predicted by Andre Roberge and Nathan Weiss~\cite{roberge1986gauge} there is
  first order phase transition for temperature above $T_{RW}$, see Fig.~\ref{fig:RWphase}.
  In general there are three regimes for $T$--$\mu_I$ plane:
  I: $T<T_c$, where $T_c$ is pseudo critical temperature at zero value of chemical potential, this is confining phase, no phase transition;
  II: $T_c<T<T_{RW}$, intermediate range, crossover between confinement and deconfinement takes place at some value $\mu_I<\pi/3$;
  III: $T>T_{RW}$, deconfining phase, Roberge-Weiss first order phase transition is located at $\mu_I = \lr{\pi+2\pi n}/3$ (here $\mu_I$ -- quark chemical potential).
  For each of this three regimes we use different fit functions for $n\lr{\mu_I}$ to describe lattice data.
  For the first and second regimes we use truncated Fourier series (one term for $T/T_c=0.84$, until seven terms for $T/T_c=1.08$).
  For the third regime we use polynomial fit $n(\mu_I)=a_1 \mu_I+a_3 \mu_I^3$.
  For all these ansatzes we restore $n$ in imaginary region and predict number density at
  real value of quark chemical potential $\mu$ (Figs.~\ref{fig:bdi} and~\ref{fig:bdr}).
  
  We use 300 $Z_n$'s ($Z_{3}$, $Z_{6}$, $\ldots$, $Z_{900}$) for $T/T_c=0.84,\; 0.93,\; 1.035,\; 1.08$ and
  about 100 $Z_n$'s for $T/T_c = 0.99,\; 1.2,\; 1.35$.

\section{Fitting of $Z_n$ at fixed $n$}

\begin{figure}[h]
    \begin{minipage}[t]{0.5\textwidth}
      \centering
      \includegraphics[width=1.\textwidth]{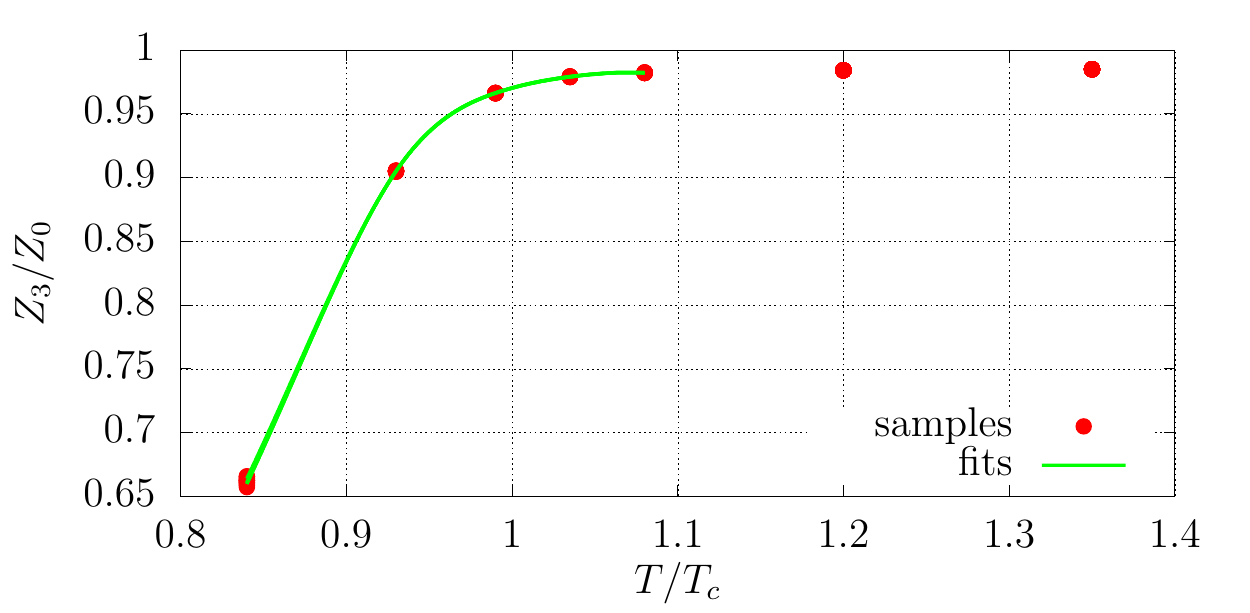}
      
      \vspace{-1.5ex}
      \includegraphics[width=1.\textwidth]{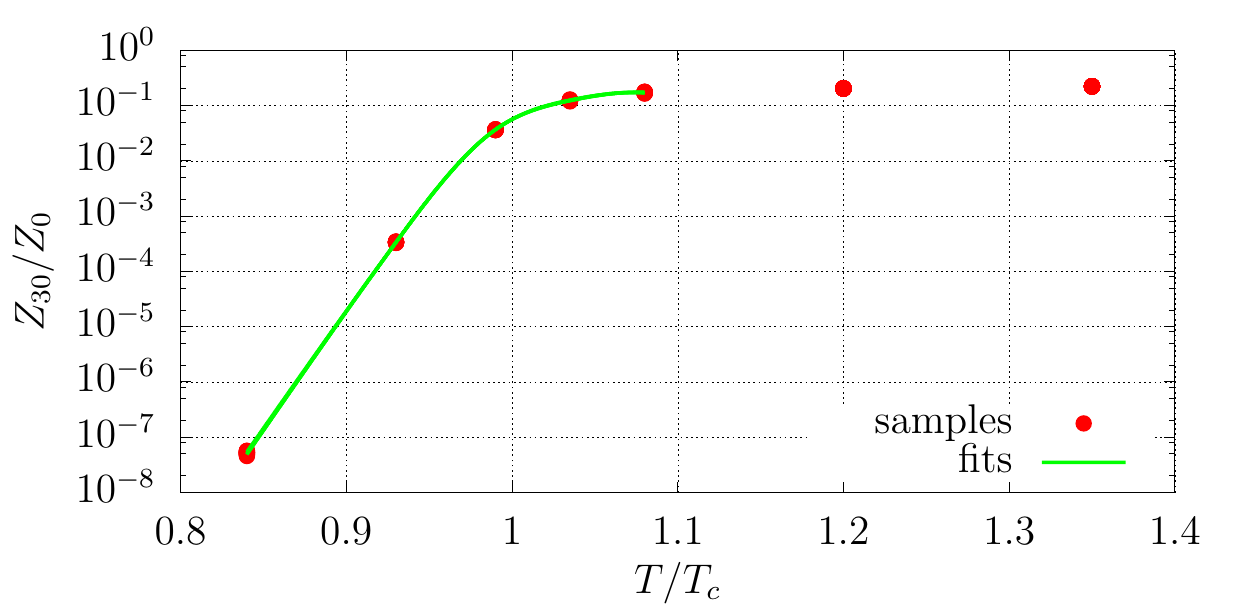}
      
      \vspace{-1.5ex}
      \includegraphics[width=1.\textwidth]{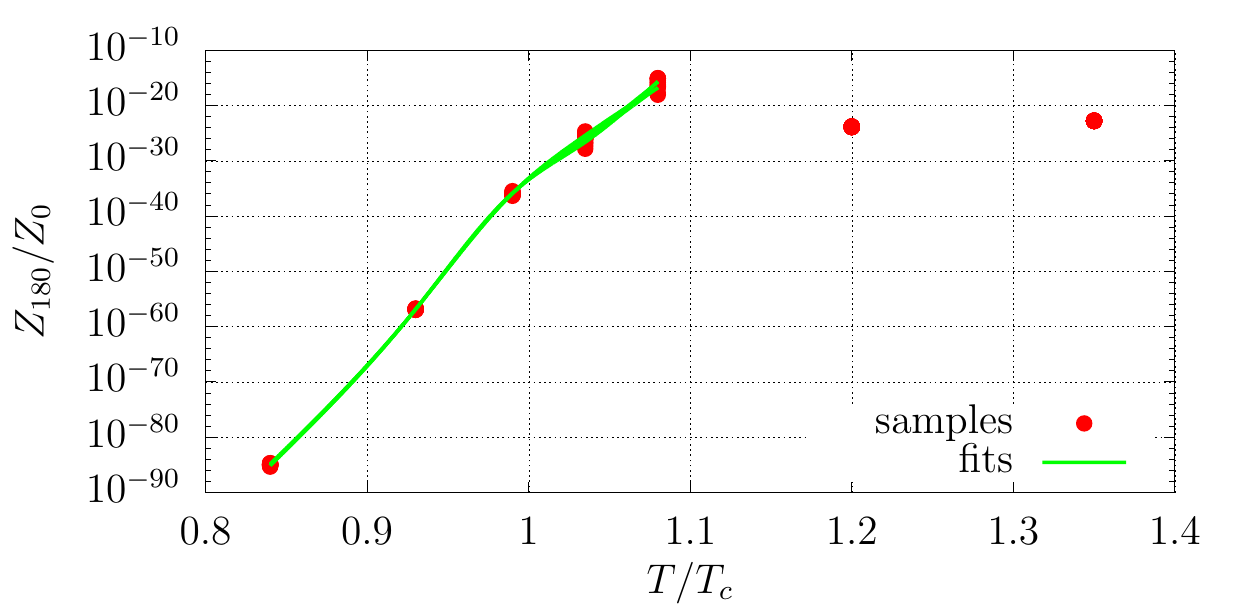}
    \end{minipage}
    \hspace{\fill}
    \begin{minipage}[t]{0.5\textwidth}
      \centering
      \includegraphics[width=1.\textwidth]{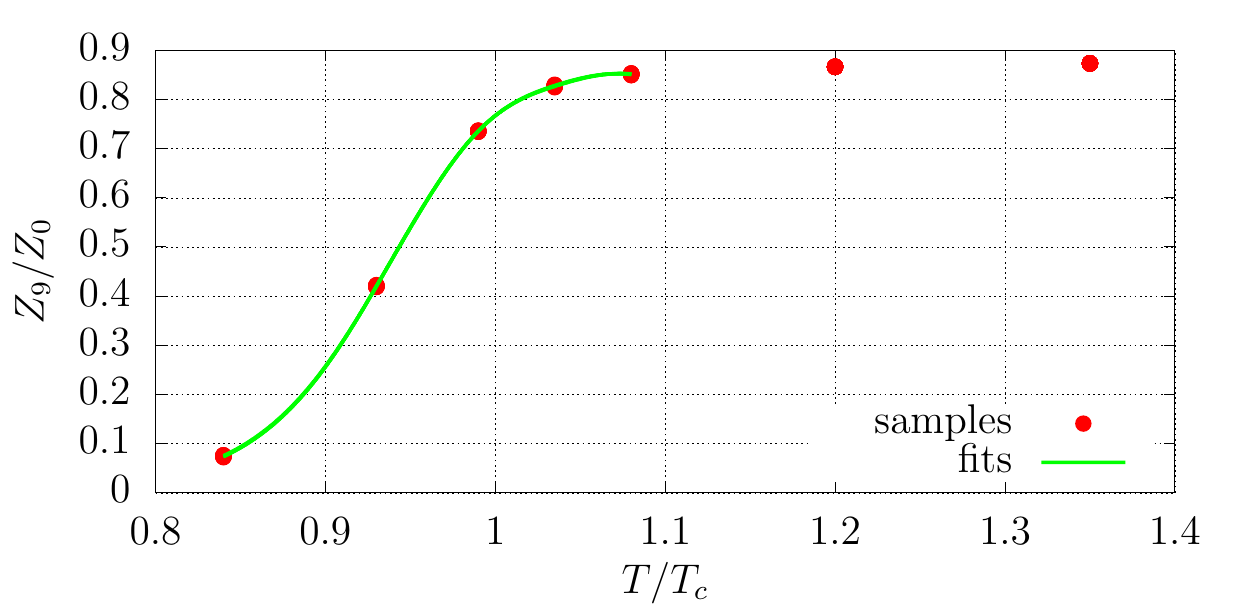}
      
      \vspace{-1.5ex}
      \includegraphics[width=1.\textwidth]{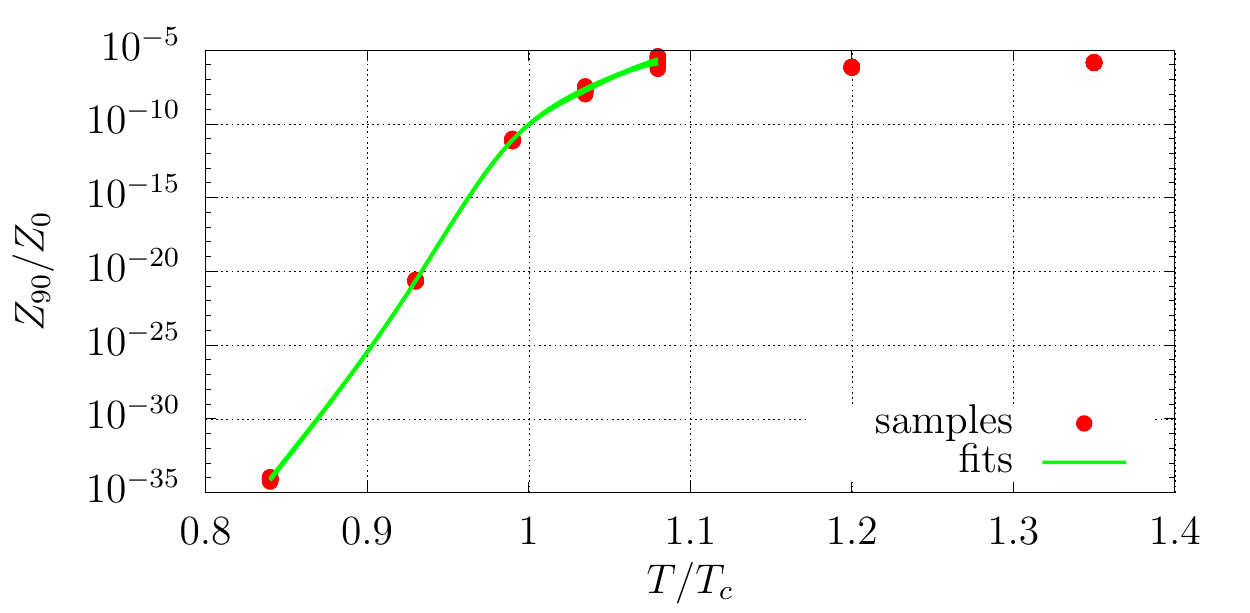}
      
      \vspace{-1.5ex}
      \includegraphics[width=1.\textwidth]{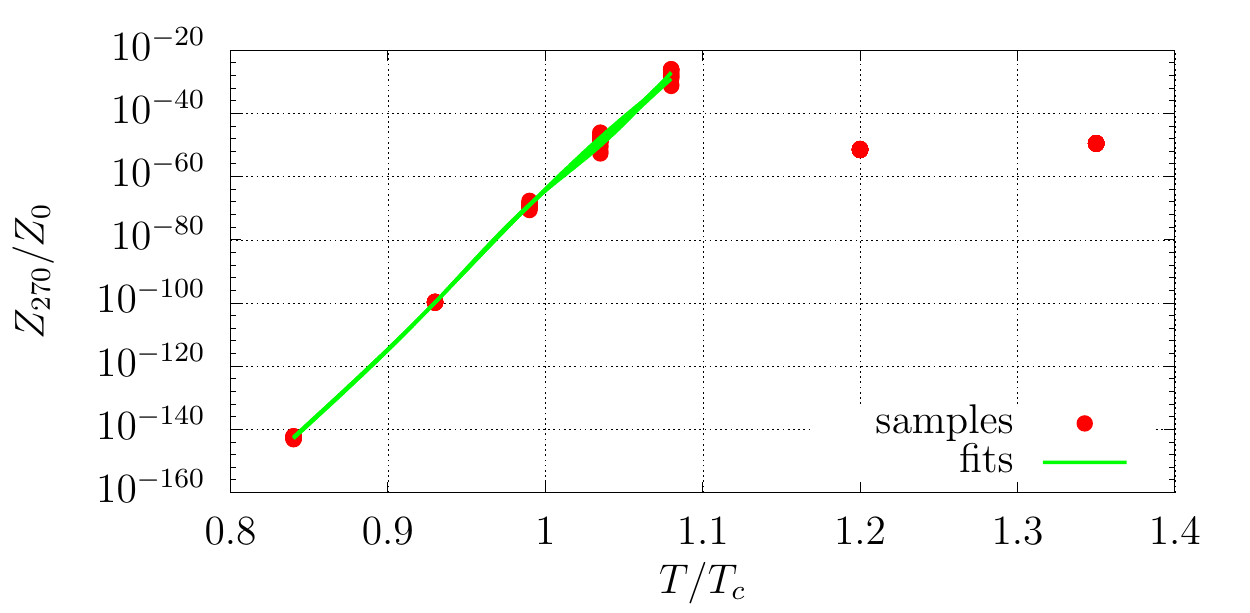}
    \end{minipage}
    \begin{center}
      \caption{Fitting procedure for different $Z_n/Z_0$. \label{fig:fitZn}}
    \end{center}
  \end{figure}

  We compute $Z_n$ at seven values of temperature. After that we can compute any observable.
  To interpolate $Z_n$ between existing temperature values we perform cubic spline for $\log Z_n$ (Fig.~\ref{fig:fitZn}) for fixed $n$.
  We are interested in temperature values near $T_c$ to restore crossover line.
  Note, that for $Z_n$ with large $n>90$ cubic spline can not be used in full range $T\in [0.84, 1.35]T_c$,
  otherwise cubic spline produces non physical dependence near $T_{RW}$.
  We fit only in range from $T/T_c=0.84$ to $T/T_c=1.08$.
  
\begin{figure}[h]
    \begin{minipage}[t]{1.\textwidth}
        \centering
        \includegraphics[width=.8\textwidth]{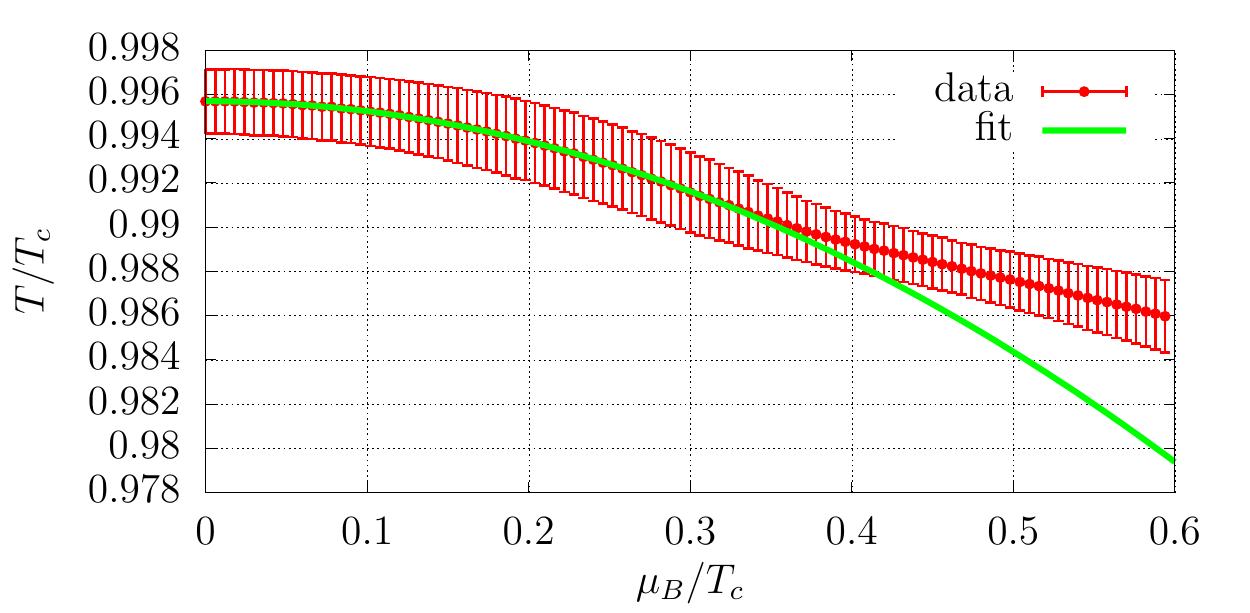}
      
        \caption{Crossover line restored via inflection point of $\chi (T)$. \label{fig:crossover}}
    \end{minipage}
  \end{figure}

  To determine crossover line we study temperature dependence of baryon density susceptibility $\chi\sim\partial^2/\partial\mu^2 \log Z_G$
  and calculate numerically $\partial\chi/\partial T$.
  For numerical computation of $\partial\chi/\partial T$ we compute 1800 temperature points in region $T/T_c=0.9\ldots 1.08$ needed for good estimation of derivative.
  Position of maximal value of $\partial\chi/\partial T$ are presented in the Fig.~\ref{fig:crossover}.
  The parametrization of crossover line is $T_c(\mu_B^2)=T_c\lr{c-\kappa\, \mu_B^2/T_c^2}$.
  In our case we use additional parameter $c = 0.9956 \pm 0.0015$, because temperature determination also has error ($\delta (T/T_c) < 5\%$).
  From our calculation we obtain $\kappa = -0.0453 \pm 0.0099$.

\section{Concluding Remarks}

  In this letter we presented evidence that the canonical approach can be useful
  to study QCD matter at non zero baryon chemical potential.
  We believe that canonical approach is one of the most promising approaches to study
  strong interacting matter from the first principles.
  Using analytic continuation for fugacity expansion we show that
  canonical partition functions can be obtained for any temperatures
  from observable computed at imaginary values of chemical potential.
  Lattice QCD at this region is free from sign problem and provide
  us with good framework to restore canonical partition functions.
  
  However, to obtain the grand canonical partition function we need to perform
  integration of baryon density.
  To accomplish this task we fit the data for baryon density to a chosen fitting function.
  But different fitting functions lead in general to different values for $Z_n$, thus introducing a systematic uncertainty.
  The fugacity expansion~(\ref{Eq:FugacityExpansion}) indicates that truncated Fourier series
  should be appropriate choice to fit the baryon density at imaginary chemical potential~\cite{DElia:2004ani}.
  Using truncated Fourier series we compute $Z_n$ for
  any temperature in the range $T/T_c=0.84\ldots 1.08$.
  
  From computed $Z_n$ we restore baryon density susceptibility and predict crossover line
  with the curvature $\kappa = -0.0453 \pm 0.0099$.
  
  As a next step we will decrease the quark mass and the lattice spacing to obtain results for more realistic parameters.
  Also we plan to study volume dependence of our results.
  
\section*{Acknowledgment}

  This work was completed due to support from Russian Science Foundation grant 15-12-20008.
  Work done by A. Nakamura on the theoretical formulation of $Z_n$
  for comparison with experiments was supported by JSPS KAKENHI
  Grant Numbers 26610072 and 15H03663.
  In computations we used Vostok-1 at FEFU and
  computational resources of Cybermedia Center of Osaka University
  through the HPCI System (ID:hp170197).

\bibliography{FDQCD}

\end{document}